\documentclass[11pt,twoside]{book}
\usepackage{konkolyproc2}
\usepackage{longtable}
\usepackage{amsmath,amssymb}
\usepackage{graphicx}
\usepackage{lscape}
\usepackage{index}
\usepackage{natbib}
\usepackage{bigdelim}
\usepackage{multirow}
\makeindex

\begin{document}

%%%%%%%%%%%%%%%%%%%%%%%%%%%%%%%%%%%%%%%%%%%%%%%%%%%%%%%%%%%%%%%%%%%%%%%%%%
\pagestyle{myheadings}
\setcounter{equation}{0}\setcounter{figure}{0}\setcounter{footnote}{0}
\setcounter{section}{0}\setcounter{table}{0}\setcounter{page}{1}
\markboth{Duffau et~al.}{Chemical abundances of distant RR Lyrae 
stars in the VSS}
\title{Detailed chemical abundances of distant RR Lyrae stars in the 
Virgo Stellar Stream}
\author{S. Duffau$^{1,2}$, L. Sbordone$^{1,2}$, A. K. Vivas$^3$, 
C. J. Hansen$^4$, M.~Zoccali$^{2,1}$, M. Catelan$^{2,1}$, D. Minniti$^5$, 
\& E. K. Grebel$^6$}
\affil{$^1$Millennium Institute of Astrophysics, Santiago, Chile\\
$^2$Instituto de Astrof\'isica, Pontificia Universidad Cat\'olica de Chile, 
 Santiago, Chile\\ 
$^3$Cerro Tololo Inter-American Observatory, La Serena, Chile \\
$^4$Dark Cosmology centre, Niels Bohr Institute, University of Copenhagen, 
Denmark \\
$^5$UNAB, Santiago, Chile \\
$^6$ARI, ZAH, Heidelberg University, Heidelberg, Germany}
%%%%%%%%%%%%%%%%%%%%%%%%%%%%%%%%%%%%%%%%%%%%%%%%%%%%%%%%%%%%%%%%%%%%%%%%%%%

\begin{abstract} We present the first detailed chemical abundances for 
distant RR~Lyrae stars members of the Virgo Stellar Stream (VSS), 
derived from X-Shooter medium-resolution spectra. Sixteen elements 
from carbon to barium have been measured in six VSS RR~Lyrae stars, 
sampling all main nucleosynthetic channels. For the first time we will 
be able to compare in detail the chemical evolution of the VSS progenitor 
with those of Local Group dwarf spheroidal galaxies (LG dSph) as well as 
the one of the smooth halo. 
\end{abstract}

\section{Project presentation}

Numerous substructures have been found so far in the halo of our galaxy, 
and although their census is far from complete, the main challenge now 
seems to be finding the connections between them, characterising their 
intrinsic properties, and identifying their origin. Following several 
photometric and spectroscopic studies, it is now clear that a large 
overdensity of stars containing several density peaks and velocity 
signatures is present in the direction of the Virgo constellation, 
its most prominent velocity feature being the VSS RR Lyrae (RRL) stellar 
stream (Duffau et~al. 2014 and references therein). In order to 
investigate its nature we started a spectroscopic campaign to obtain 
medium-resolution spectra of some of its constituent RRL members to 
obtain the first detailed abundance analysis of this kind performed on 
an RRL stellar stream. RRL stars have proven to be excellent tracers 
of halo substructures, due to the quality of the distances that can be 
derived from them, and the relatively straightforward techniques that 
lead to their identification. However, they have so far never been used 
to determine the detailed chemistry of the substructures they reveal, 
since detailed abundance analysis is difficult for these pulsating 
variables. This is unfortunate, since the crucial chemical assessment 
of the substructures must then be derived from other stellar populations 
whose membership in the same substructure is less certain.

The VSS is located at $\sim$20 kpc from the Sun and its core is at 
RA $\sim 187^\circ$ and Dec $\sim -1^\circ$. We have selected 7 targets 
in the area and obtained X-Shooter@VLT spectra at a high signal to noise 
(80 or higher) and a resolution of $\sim$6000 for the UVB arm and 
$\sim$10000 for the VIS arm, covering altogether approximately from 
2900 to 10000 \AA. The data were obtained in service mode. In addition 
to our target stars, we obtained spectra for two comparison RRL stars 
from the For et~al. (2011) sample as well.
Six of the targets had been found in a coherent structure in phase-space 
(heliocentric distance - radial velocity space) using a group finding 
algorithm developed in Duffau et~al. (2014) and one is a kinematical 
member from the first spectroscopic study in Duffau et~al. (2006). 
All RRL stars selected are of type ab, which is more easily identified 
and less affected by contaminants than the type c members. 

\section{Preliminary results}

We have conducted a first 1D LTE atmospheric parameters determination 
and chemical analysis using the MyGIsFOS code (Sbordone et~al. 2014). 
An example of a spectrum section is shown in Figure~1. We found numerous 
Fe\,I and Fe\,II lines allowing for the determination of the stellar 
parameters directly from the spectra. In only one case we had to fix 
the effective temperature to determine the abundances. 

\begin{figure}[!h]
\includegraphics[width=1.0\textwidth]{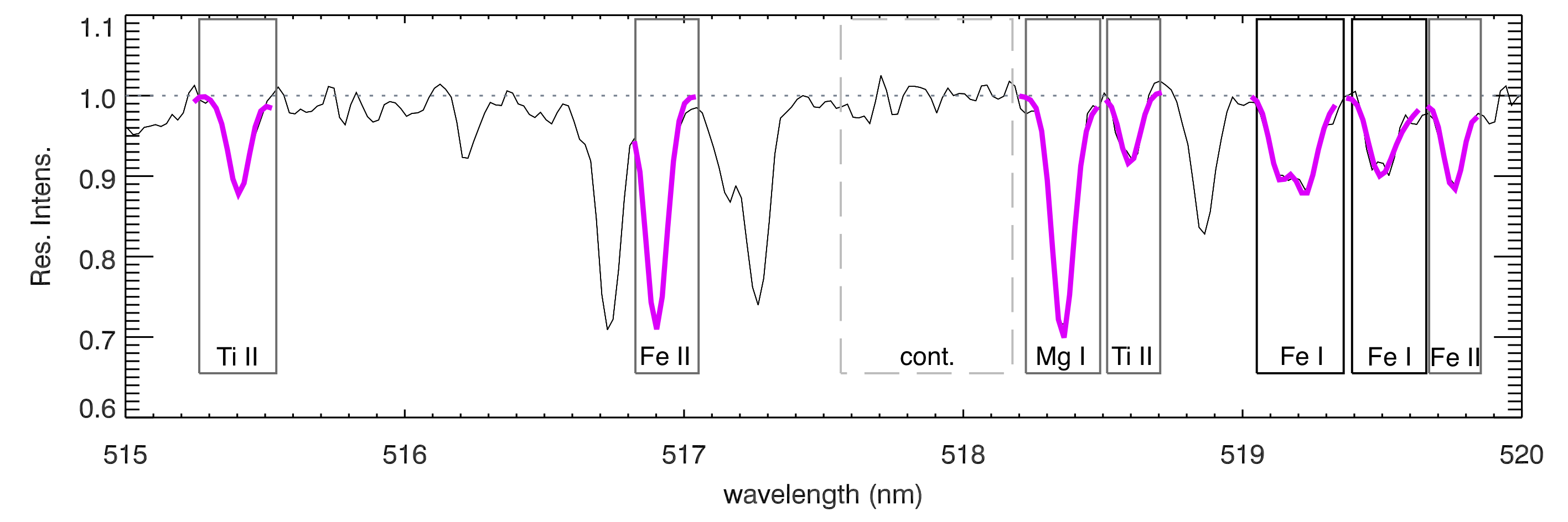}
%%%\vspace*{-2mm}
\caption{Sample section of spectra (thin line) displaying some of 
the selected line and continuum features defined for analysis 
(solid and dashed line boxes), as well as the best fits for abundance 
determination using MyGIsFOS (thick line within box) and the continuum 
(dotted horizontal line). The elements under study are marked in each box.} 
%%\label{authorsurname-fig1} 
\end{figure}

The observations were performed attempting to constrain the phase 
range away from the rising branch of the light curve to where the 
effective temperatures are cooler and the abundance analysis is more 
reliable. Regrettably, the ephemerides we originally used to plan the 
observations proved outdated and the observations ended up spanning 
a larger range in phase than intended. As a result, one of our target 
stars could not be analysed due to unfavourable phase (at 0.97) 
coupled with a very low metallicity. Of the remaining six targets, 
four lie close or within the safest phase range (at 0.25, 0.34, 0.51, 
and 0.82), and parameters and abundances could be obtained for them, 
while two lie outside of it (at 0.10 and 0.14) and a few of the lines 
of some of the elements were lost. 

\begin{figure}[!h]
\includegraphics[width=1.0\textwidth]{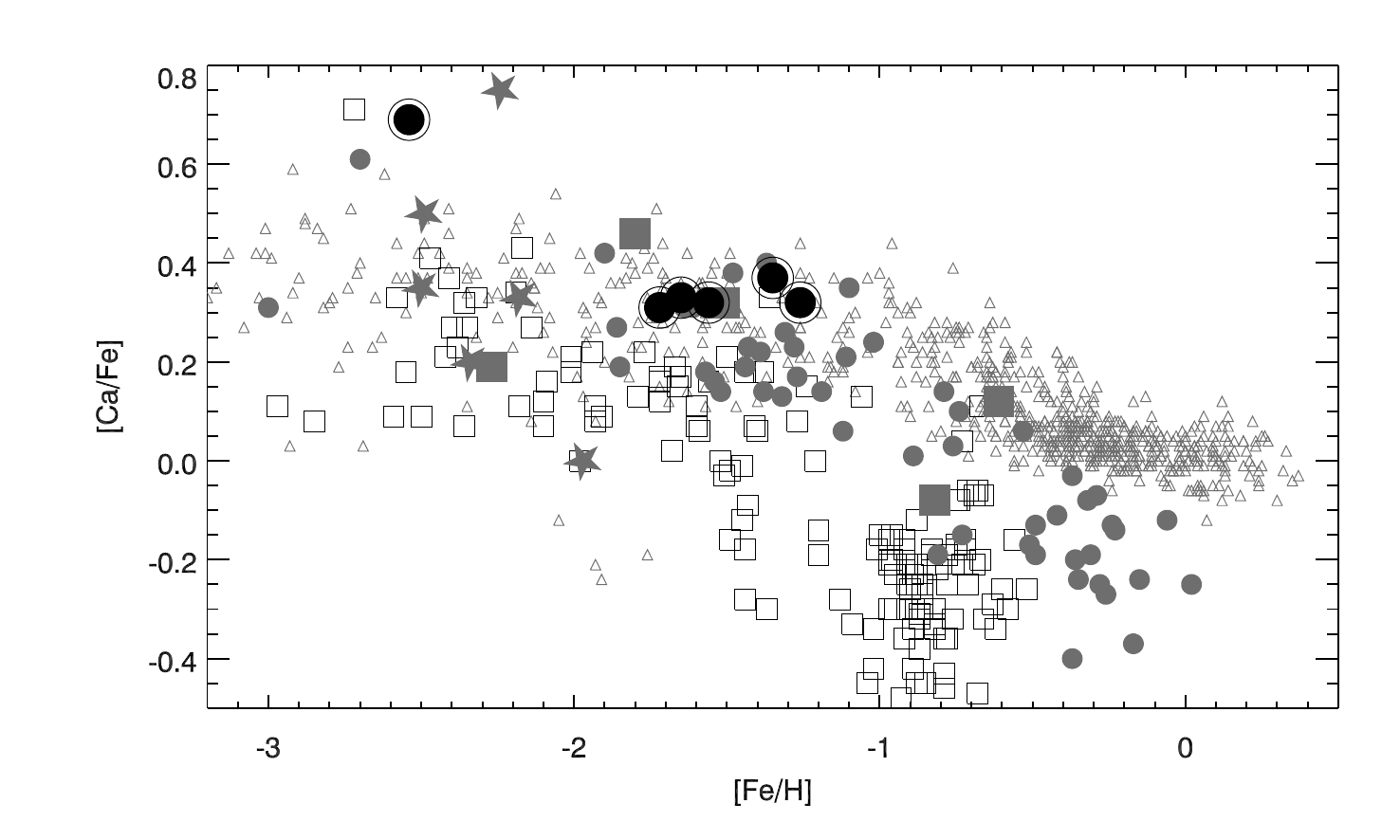}
\vspace*{-4mm}
\caption{[Ca/Fe] ratio as a proxy for the $\alpha$-abundance of the VSS 
sample (large circles) as compared to LG dSph galaxy data (Classical; 
open squares, Sgr; small circles), MC RRL (stars), and Milky Way data 
(open triangles).} 
%%\label{authorsurname-fig1} 
\end{figure}

%%%\vspace*{-6mm}

\begin{figure}[!h]
\includegraphics[width=1.0\textwidth]{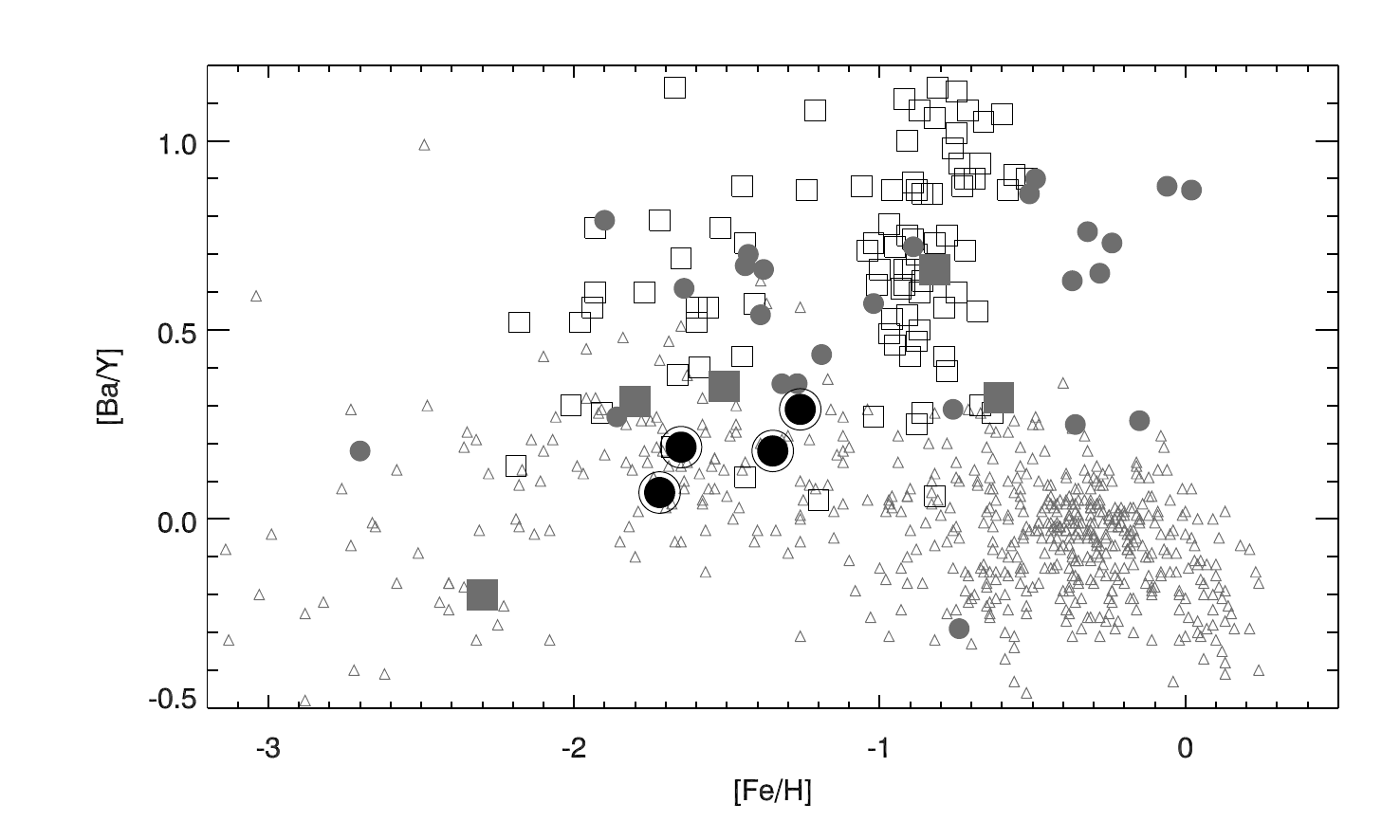}
\vspace*{-4mm}
\caption{[Ba/Y] ratio for the VSS sample, same symbols as Fig.~2.} 
%%\label{authorsurname-fig1} 
\end{figure}

Figure 2 shows the results for the six RRLs in our sample and their 
measured calcium abundance versus metallicity. The calcium abundance 
is taken here as a representative of the $\alpha$-elements abundance of 
the system. The value for the observed stars is compared with those 
for a selection of ``classical'' dwarf spheroidal galaxies (dSph; 
including Fornax, Sculptor, Carina, Leo, Ursa Minor, Sextans and Draco), 
for stars and clusters belonging to the Sagittarius dwarf spheroidal 
galaxy (Sgr dSph; including Terzan 8, Arp 2, M54, Terzan 7 and Palomar 12), 
for a sample of Milky Way stars from the literature (see Sbordone et~al. 
2015 and references therein), and for six RRL stars in the Magellanic 
Clouds (MC; Haschke et~al. 2012). 

Figure 3 displays the results for four of our RRL stars that had both 
barium and yttrium measured. For the other two only one of those elements 
was measurable. The barium over yttrium ratio as a function of metallicity 
is compared to the value for the same systems as in Fig.~2, except for 
the MC RRLs whose data for these elements was not available. In the case 
of the calcium abundance, we see that the value for the bulk of the VSS 
stars compares better to larger systems like the Sgr dSph and the Milky 
Way sample, than to the lower-mass classical dSph. This suggests a 
progenitor with a high star formation efficiency, hence likely a massive 
one, as massive as the Sgr dSph or maybe even larger. In the case of the 
barium over yttrium ratio, this ratio is taken as an indication of the 
efficiency of the s-process of the system under study. It is an 
observational fact that this ratio remains high for both the Sgr 
dSph and the classical dSph galaxies when compared to the Milky Way 
sample in the metallicity range covered by the VSS targets. In this 
case the value for the VSS is more compatible with the Milky Way sample 
than with any other sample, again probably also indicating a massive 
progenitor. 

To confirm our results, we will add the information from more elements 
and other nucleosynthetic channels and study the impact of NLTE/3D 
effects in our preliminary conclusions. Up until now, these preliminary 
results show, first of all, that X-Shooter is a very effective instrument 
for determining detailed chemistry of distant halo RRL stars, as far as 
20 kpc. They also suggest that we might have found a case of an accretion 
event as significant as the Sgr dSph system, that has contributed stars 
to the Galaxy with the typical chemistry of the smooth halo component.

\section*{Acknowledgments}
\vspace*{-2mm}
This project is supported by the Ministry for the Economy, Development, 
and Tourism's Iniciativa Cient\'ifica Milenio through grant IC\,120009, 
awarded to the Millennium Institute of Astrophysics; by CONICYT's PCI 
program through grant DPI\,20140066; by Proyecto Fondecyt 
Regular \#1141141; and by Proyecto Basal  PFB-06/2007.


\begin{thebibliography}{}      

\bibitem[Duffau et~al. (2006)]{D06}
Duffau, S., Zinn, R., Vivas, A.~K., et~al. 2006, \apj, 636, 97
\bibitem[Duffau et~al. (2014)]{D14}
Duffau, S., Vivas, A.~K., Zinn, R., et~al. 2014, A\&A, 566, A118
\bibitem[For et~al. (2011)]{F11}
For, B.-Q., Sneden, C., Preston, G. 2011, \apjs, 197, 29
\bibitem[Haschke et~al. (2012)]{H12}
Haschke, R., Grebel, A. K., Frebel, A., et~al. 2012, \aj, 144, 88
\bibitem[Sbordone et~al. (2014)]{S14}
Sbordone, L., Caffau, E., Bonifacio, P., et~al. 2014, A\&A, 564, A109
\bibitem[Sbordone et~al. (2015)]{S15}
Sbordone, L., Monaco, L., Moni Bidin, C., et~al. 2015, A\&A, 579, A104

\end{thebibliography}
\end{document}